# HIERARCHICAL DISTRIBUTED PREDICTIVE CONTROL APPROACH FOR MICROGRIDS ENERGY MANAGEMENT

## Le Anh Dao*, Alireza Dehghani-Pilehvarani*, Achilleas Markou[†] and Luca Ferrarini*

*Dipartimento di Elettronica, Informazione e Bioingegneria, Politecnico di Milano, P.za L. da Vinci 32, 20133 Milano, Italy.*

*† Smart Grids Research Unit, School of Electrical and Computer Engineering, National Technical University of Athens, 15780 Zografou, Greece.*

*Abstract*— This paper addresses the problem of management and coordination of energy resources in a typical microgrid, including smart buildings as flexible loads, energy storages and renewables. The overall goal is to provide a comprehensive and innovative framework to maximize the overall benefit, still accounting for possible requests to change the load profile coming from the grid and leaving every single building or user to balance between servicing those requests and satisfying his own comfort levels. The user involvement in the decision-making process is granted by a management and control solution exploiting an innovative distributed model predictive control approach with coordination. In addition, also a hierarchical structure is proposed, to integrate the distributed MPC user-side with the microgrid control, also implemented with an MPC technique. The proposed overall approach has been implemented and tested in several experiments in the laboratory facility for distributed energy systems (Smart RUE) at NTUA, Athens, Greece. Simulation analysis and results complement the testing, showing the accuracy and the potential of the method, also in the perspective of implementation.

## I. INTRODUCTION

It is well known that the oscillations of fossil fuel price and the environmental concerns of their widespread usage have raised the interest in alternative energy resources, including renewables and storages, possibly managed through optimal control strategies. Besides being cleaner, these energy resources can often be placed in the vicinity of the end users, thus reducing the energy losses related to electricity transmission. This entails a radical change in the structure of the energy system, where the new electricity network scenario includes many small and distributed generators connected to medium and low voltage grids, as opposed to a few large generators connected to high voltage. The concept of microgrid appears to be a promising solution to properly address this type of scenario.

Accordingly, many research activities flourished recently around these scenarios, and the development of optimal control solutions is one of them. The typical attacked problems include, for example, economical management of the resources to trade with the energy market [1–3], reduction of greenhouse gas emission [4], tracking day-ahead committed power profile with the distribution network, in presence of various disturbances and uncertainties and with the general goal to preserve a given level of use comfort. Solutions employ different techniques such as heuristics [5], genetic algorithms [6], game theories [7–9] and optimization techniques [4], [10]. More specifically, some research lines attack the problem of wisely operating (B)ESS (Battery Energy Storage System) and DERs (Distributed Energy Resources) and modifying pre-scheduled consumption profile of flexible loads, possibly involving end-users in the decision-making process. In particular, [2,11–13] and [14] discussed the role of the (flexible) load in supporting the grid ancillary services and frequency regulation. Others, like [15], [16], focus on economic or operation optimization of microgrid.

A particularly useful methodology in the context of microgrid management is Model Predictive Control

(MPC), which is well suited to deal with a large amount of constraints of different types that have to be imposed in real time in microgrids. This technique has been exploited for example in [2,3,11,12] and [17–23]. In [2], [3] and [21] the flexible loads are modelled as a predefined range of possible load consumption and the microgrid controller can directly determine a load profile as long as it fulfils the range. This approach neglects the dynamic of load flexibility in the shiftable loads category (e.g., heating/cooling system, refrigerator, washing machine, etc.) since the size of the predefined range in one step of this load type can not be fixed in advance and it depends on the value of decision variable in the previous steps. On the other hand, in [17] and [23], the load side may reveal its model which later will be formulated in the optimization problem in a centralized controller.

Concerning the control scheme, a centralized predictive scheme is considered in [2], [3], [12], [17] and [18]. However, it is well-known that this scheme presents issues of scalability, computational burden, failure of single unit, adaptability, etc. Recent works are putting more attention to the distributed MPC and hierarchical control schemes, such as [11], [19]. In particular, in [19], a two-layer control scheme based MPC operating at two different timescales has been studied. In the paper, some details on the markets (e.g., imbalance charge, difference in purchasing and selling tariffs) are neglected and the flexible load is not considered. On the other hand, the paper [11] employs a sequential distributed MPC on energy management problems in the microgrid; however, some details are missing, including the presence of RES (Renewable Energy Resource), the difference in purchasing and selling tariffs, and the difference in ESS charge and discharge efficiencies. These assumptions lead to the formulated MPC problem in each local controller rather simple instead of using Mixed-Integer Quadratic Programming (MIQP) as [2] or splitting an original variable into multiple variables as in [3]. Also, the research focuses on planning level with 1-hour sampling time. Focusing on distributed MPC (DMPC) applied to the building control problem, [24] has discussed a non-iterative, non-cooperative DMPC algorithm based on the classification discussed in [25]; in the meanwhile the algorithms in [26], [27], [28] belong to class of distributed optimization for the centralized optimization problem employing different methods such as the proximal Jacobian alternating direction method of multipliers (ADMM) in [26], the primal-dual active set in [27], the Lagrangian dual method in [28]. In the same class of distributed optimization for the centralized optimization problem, the incremental proximal method is discussed in [29]; the method is generally believed as a more stable approach than the gradient-based one.

The present paper addresses the problem of management and coordination of energy resources in a typical microgrid, including flexible loads, energy storages and renewables. In the present paper, an innovative control scheme is proposed, based on a suitable combination of hierarchical and distributed model predictive control (H-DMPC), in order to perform economical management of power flows in a microgrid in grid-connected mode, equipped with RES, an ESS and flexible loads, that in our modeling metaphore is represented by a group of smart buildings where the thermal comfort is provided by electrical equipment. Basically, the overall microgrid is separated into two layers, naming microgrid level and load (or building) level. Any forecast errors are mitigated, compensated taking advance from small sampling time of the negotiation between two levels with respect to the time scale of the main target factors in the microgrid operation. The microgrid level manages the ESS and RES to track day-ahead power profile and maximize the energy profit through trading electricity activities. To do so, a standard MPC problem is formulated employing shrinking horizon and MIQP. The latter one, on the contrary employs an innovative distributed, shrinking horizon, model predictive control technique with coordination, to achieve consensus solution for coupled control variables among users/loads/buildings regarding comfort and cost optimization. The technique employs, in MPC framework, iterative, proximal minimization based algorithms, where the local objective functions of users are refined by adding penalty terms which penalize a consensus residual factor (see in [29] [30] and references therein). Specifically, the structure of incremental proximal algorithms in [29] is used in our paper which allows the loads in our consideration having a maximum autonomy within the microgrid. The technique gives the possibility not to share any information except values of the coupled variables; direct connections between involved users are not necessary but all the exchanged information is collected and evaluated together at every iteration in a central coordination unit which helps to reduce the communication burden and privacy. Low computational load and small memory required are other advantages coming from the considered technique.

A few papers including [31], [32] and [33] have been studying the applications of this technique in the context energy management of smart grid and microgrids. Beside the fundamental differences in the control architecture, control problem, as well as some technical details in the ESS models, electricity tariffs, imbalance charge, our paper introduces an additional degree of freedom in the central coordination unit to evaluate the exchanged information (or solution) collected from each agent, while the local cost function of each agent remains unchanged and the weight information is unknown by any agents. In detailed, an optimization problem which is a weighted sum of quadratic terms associated with the exchanged information is considered; the weights are selected based on a common rule between the users (e.g., high weight is assigned to the user which contributes significantly or more trustworthy to the overall system and vice versa). The consideration implies that the exchanged information of an agent which has high weight with respect to others' will get more attention in the evaluation of the central coordination unit.

Our research largely extends the studies presented in [3], [20] and [34] which discusses a basic microgrid energy management scenario in a centralized model predictive control framework. Moreover, the validity of our approach has been investigated and tested in both simulation and laboratory environments.

The rest of the paper is organized as follows. The microgrid setting and considered scenarios are described in Section II, the corresponding model, control architecture and algorithms are introduced in Section III. The simulation results and experimental results are illustrated and discussed in Section IV and Section V, respectively, before the concluding remarks reported in Section VI.

## II. CONTROL SETTING OVERVIEW

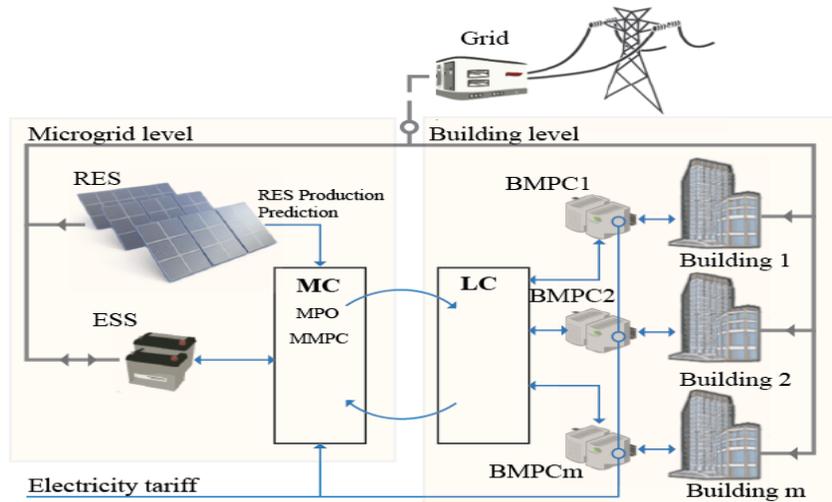

Fig. 1. Microgrid setting and designed control system (Power flow: gray, bold line; Information, control signal: blue line)

The microgrid considered in this paper is sketched in Fig. 1. It comprises a group of smart buildings which can vary from small to large loads (e.g., residential buildings, airports, shopping districts and commercial buildings), a RES (e.g., wind turbines or photovoltaic panels) and an ESS (Energy Storage System).

All components of the microgrid are connected on the same electricity bus and linked with the main grid by a Point of Common Coupling (PCC), assumed to be closed (grid-connected mode). The islanded mode is not considered here, since one of the objectives is the maximization of the economic benefit related to electricity trading. The electricity required to supply the load side can be taken from either the RES or the ESS, or purchased from the grid. The electricity in excess generated by the RES can be either stored in the ESS or sold to the grid. Smart buildings are here endowed with some flexibility in changing the load demand to pursue the incentive received from changing its planned consumption. Such a flexibility is enforced eventually through a local smart control system, which takes again the form of an MPC, which is however out of the scope of the present paper.

*A. Overall control architecture*

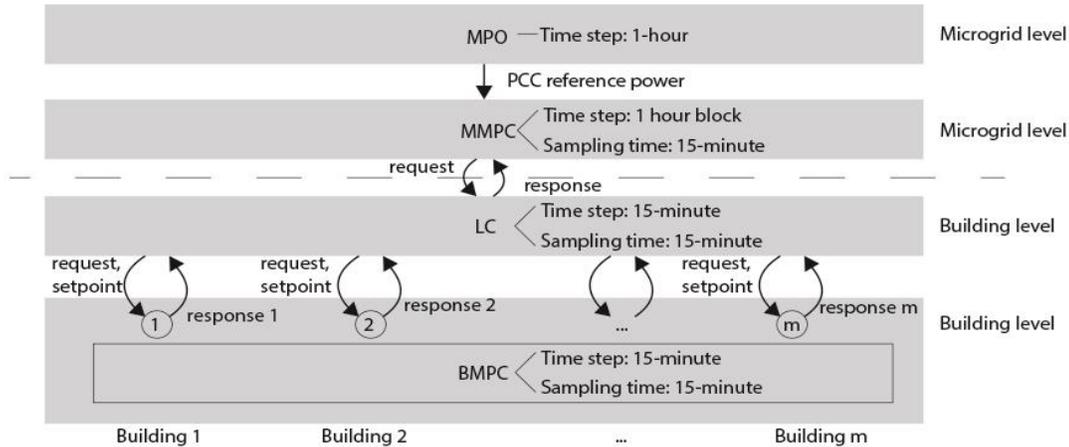

Fig. 2. Overall control architecture of the microgrid

As shown in Fig. 1 and Fig. 2, the microgrid energy management is separated into two problems, each solved by two different controllers. They are the Microgrid Coordinator (MC) and Load Coordinator (LC) and are working with the same timescale, by default set to 15 minutes. Notice that, in Fig. 2, the time step term refer to duration of a discrete time step while the sampling time term implies rate of running of related entities.

The MC is in charge of fulfilling as much as possible the promised hourly profile exchanged with the grid, in front of uncertainties of future power production of RES as well as load power consumption and other disturbances (e.g., weather conditions). Basically, the MC consists of a controller and an optimizer. Its controller part decides for the battery (i.e. ESS) charging and discharging through an MPC-based approach, denoted as MMPC, that receives as inputs the predicted production of the RES, the load demand, the time-of-use (TOU) electricity tariffs, the power profile promised to the grid, and aims at the minimization of the total energy cost of the microgrid, possibly asking for a change of load profile. On the other hand, microgrid planning optimizer (MPO) is the optimizer part of the MC which is designed to generate references for hourly day-ahead power exchanged through PCC (or for short reference power) considering as input the prediction of day-ahead RES generated power and load consumption. This optimizer is executed once a day with the prediction horizon of 24-hour and 1-hour sampling time which matches to the requirement of the day-ahead market. Finally, the MC also takes a responsibility of communicating to the LC a load change request (or request) for changing their planned consumption.

Although the sampling time of the MMPC is 15 minutes, the reference power is given as an hourly profile and its prediction horizon is based on a 1-hour block which causes the variation of the duration of the first 1-hour block of the prediction horizon from 15 minutes to 60 minutes. Notice that the prediction horizon of the MMPC setting is a fixed value as in the standard MPC approach.

On the other side, the LC receives the request from the MC, and initiates and coordinates an iterative process among buildings to try to fulfil the requests and minimizes costs.

In detail, the LC behaves as an aggregator that (i) initially receives the request to change load profile from the MC and then dispatches this information to the load controllers, (ii) receives the response from load controllers, (iii) computes (optimally) and distributes new load power profiles, iterates until an agreement is reached. More specifically, each building is equipped with an MPC-based controller for negotiation (denoted as BMPC) that works with the same sampling time as the MMPC, but the formulation is based on a 15-minute block. This difference in both the sampling time and the length of each time step generates a special setting of the prediction horizon (denoted as dynamic shrinking horizon) of the BMPC as depicted in Fig. 3. While a normal MPC rolls the prediction horizon over time steps with a fixed length of the horizon, in the dynamic shrinking horizon the prediction horizon is changed while rolling. As for communication between Microgrid and Load level, at the beginning of each time instant, the BMPC receives the request (i.e., maximum change of total local consumption) from microgrid. Then, following the iterative distributed MPC with a coordinator framework, the level of change for each of them is established by way of an interaction and negotiation process among the local controllers. For the sake of simplicity, only the planning level (or high-level controller BMPC) is here discussed for the load controller while the low-level control system is not included. However, some level of disturbances for the load side is still presented by assuming that the real consumption of the buildings equals to the planned load consumption plus white noises.

The process of solving the MPO, MMPC and exchanging information between the MC and the LC is named as microgrid negotiation phase, while we denote user negotiation phase as the process of solving the BMPCs and exchanging information between the LC and the BMPCs. Notice that only one cycle of negotiation between the MC and the LC is studied in this paper, while multiple cycles consideration is instead a subject of future work. The communication between these units are depicted also in Fig. 4.

After the negotiation phases (i.e., microgrid negotiation phase and user negotiation phase), which take a few seconds or a minute in both simulation and laboratory environment, the control phase takes place for both the BMPC and the MMPC for the rest of the sampling time.

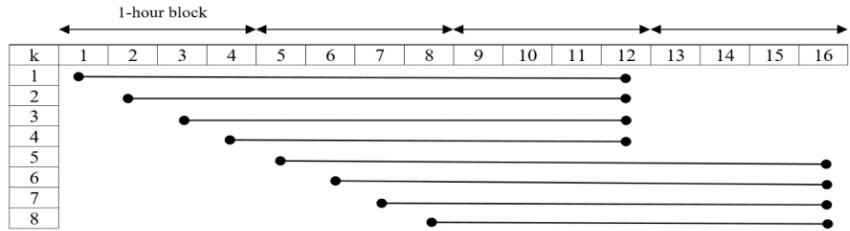

Fig. 3. Prediction horizon in dynamic shrinking horizon MPC

### B. Control objectives

This section provides an overview on the objectives, both technical and economic ones, imposed to the controllers at the microgrid and building levels.

For the main control objectives in MMPC, a similar approach as the one in [3] has been here developed, including energy profit maximization, minimization of imbalance charges and smoothness of the power flow terms. As for the MPO, only the first and third terms in the above control objectives are considered. As for the BMPC, the main control objective is considered as follows:

- *Energy cost minimization* – Buildings exploit variation of the market tariffs to consume energy in such a way that minimizes the cost for heating system in the building.

- *Incentive maximization* – The load power profile can be modified in the interest of the entire microgrid regarding the incentive. The higher the change, the greater incentive load side will receive for the compensation.

- *User comfort maximization* – Even though the changing of the desired load consumption may result in the incentive maximization and energy cost minimization for the users, it basically brings the actual temperature of the building away from the setpoint. Therefore, the end-users should also track their setpoint to maximize their comfort.

In the following, the design of the microgrid coordinator (MC, Section 3) and of the load coordinator and building controllers (LC and BMPC, Section 4) are given.

## III. Design of the Microgrid Coordinator (MC)

This section is devoted to the microgrid level to develop microgrid planning optimizer (i.e., MPO) and microgrid MPC (i.e., MMPC) and to clarify their functionalities. At the end, the optimization problems are formulated as MIQP ones which are solved by using ILOG's CPLEX 12.8 (an efficient solver based on branch-and-cut algorithm).

### A. Nomenclature

The main parameters and decision variables used in this section are described in Table I, where, for simplicity, the subscript $i$ is omitted when referring to the $i$-th step ahead in the prediction (or control) horizon.

Table I. Main nomenclature in the microgrid level

| | |
|---|---|
| $k$ | Discrete time step based on 1-hour block |
| $N_M$ | Prediction and control horizon of the MMPC |
| $N_{MPO}$ | Prediction and control horizon of the MPO |
| $c_k^b, c_k^s$ | Purchasing tariff and selling tariffs respectively [Cents /kWh] |
| $\eta_d, \eta_c$ | The discharge and charge efficiencies of the ESS ($0 \leq \eta_d, \eta_c \leq 1$) |
| PEN | Fixed and prefined value penalty tariff for imbalance charge [Cents /kWh] |
| | |
| $P_k^g$ | Exchanged power flow at the PCC (or PCC power for short), $P_k^g \geq 0$ and $P_k^g < 0$ denote purchased and sold electricity with the grid respectively [W] |
| $P_k^b$ | Power supplied by the ESS, non-negative values for charging and negative values for discharging [W] |
| $P_k^R$ | Power produced by RES, non-negative values for producing [W] |
| $P_k^l$ | Electrical power absorbed by the load, non-negative values for consuming electricity [W] |
| $E_k^b$ | Energy stored in the ESS [Wh] |
| $P^{ref}$ | Reference power for trading electricity with the grid |
| $P_k^{exe}$ | Average of power already flowed through PCC sampled every 15 minutes within the $k$-th 1-hour block |
| $P_k^{g,sp}$ | Setpoint for the PCC power $P_k^g$ to track. |
| $P_k^{g,plan}$ | Average planned PCC power for the remaining period of the $k$-th 1-hour block [W] |
| $\bar{R}_k$ | Average of the request power sent to the LC for the remaining period of the $k$-th 1-hour block [W] |
| | |
| $T_k^s$ | The duration of the remaining time of the $k$-th 1-hour block [Hour] |
| $\delta_k^b$ | Additive binary variable: charging (1)/discharging (0) mode for the ESS, |
| $z_k^b$ | Auxiliary variable of the ESS |
| $\delta_k^c$ | Additive binary variables: Exporting (0)/Importing (1) electricity to/from the grid from/to microgrid level |
| $C_k^c$ | Auxiliary variable for electricity trading between microgrid level and the grid |
| $\delta_k^P$ | Additive binary variables: PCC power is lower (0)/ higher (1) than PCC reference power (at time instant k) |
| $R_k^P$ | Auxiliary variable for tracking PCC reference power |
| | |
| $P_{max}^g, P_{min}^g$ | Upper and lower bounds on power exchange with grid, respectively [W] |
| $P_{max}^b, P_{min}^b$ | Upper and lower bounds of ESS power [W] |
| $E_{min}, E_{max}$ | Upper and lower bounds of ESS energy [Wh] |

*B. Power balance equation and ESS model*

The power flow in the microgrid is described by the power balance equation:

$$P_k^g = P_k^b - P_k^R + P_k^l \tag{1}$$

As for ESS, we consider constant, different charge and discharge efficiencies and its energy equation as follows:

$$E_{k+1}^b = E_k^b + \eta_k.P_k^b.T_k^s, \qquad \eta_k = \begin{cases} \eta_c & if\ P_k^b \geq 0, \qquad 0 \leq \eta_c \leq 1 \\ \dfrac{1}{\eta_d} & if\ P_k^b < 0, \qquad 0 \leq \eta_d \leq 1 \end{cases} \tag{2}$$

As discussed in Section II, the remaining time duration of the first 1-hour block in the MMPC is changing and $T_k^s$ will take a value of 1, 0.75, 0.5 or 0.25 hour (that is, the model is updated every 15 minutes). However, the value of $T_k^s$ is 1h for the following 1-hour blocks.

The ESS model contains an *if-then* condition which typically requires the use of the MIQP for solving optimization problem as introduced in [35] and applied in the microgrid optimization context in [2]. As an example, we follows the same discussion as in [2] (Appendix C and equations (1-6, 18, 19)) to formulate the logical statement for the ESS model. At the end, an additive binary variable $\delta_k^b$ (denoted as $\delta$ in general case) and an auxiliary variable $z_k^b = \delta_k^b P_k^b$ (denoted $z_k^b$, $P_k^b$ as $y$, $f$ in general case) are introduced, then the original ESS model will be presented as:

$$E_{k+1}^b = E_k^b + \left(\eta_c - \frac{1}{\eta_d}\right).z_k^b.T_k^s + \frac{1}{\eta_d}.P_k^b.T_k^s \tag{3}$$

and additional constraints in the standard form of:

$$F_1\delta + F_2 y \leq F_3 f + F_4, \tag{4}$$

where $F_1, F_2, F_3$, and $F_4$ are the suitable coefficient vectors. For the additional constraints with related to the ESS model, the constraints can be presented as:

$$F_1^b \delta_k^b + F_2^b z_k^b \leq F_3^b P_k^b + F_4^b \tag{5}$$

Eq. 5 discussed the additional constraints for the ESS model for only single discrete time step $k$, the same formulation is obtained if we consider the model in MPC framework. In that case, $\delta_k^b$, $z_k^b$ and $P_k^b$ are replaced by $\delta_{k,i}^b$, $z_{k,i}^b$, $P_{k,i}^b$ in Eq. 5 where the subscript $k, i$ represents the estimation of the related terms for time instant $k+i$ evaluated at time $k$ (i.e., $i$-th time instant in the prediction horizon). The same description for any term with the subscript $k$, $i$ are expected throughout this paper. At any discrete time step $k$, $\delta^b$, $z^b$ and $P^b$ are denoted as vectors of $\delta_{k,i}^b$, $z_{k,i}^b$, $P_{k,i}^b$, respectively.

Concerning the future disturbance terms, the load power prediction is available for the MMPC receiving information from the users at the beginning of every time instant. However, this information is not available for a longer horizon (e.g., medium term, a day) of the MPO. For the sake of simplicity, in our simulations and experiments, we assume a day-ahead load consumption in the MPO as a constant profile that satifies all the buildings to perfectly follow their temperature setpoints. As for the RES production prediction, a simple predictor, denoted as persistent predictor, is chosen to test in our study for both short-term prediction in the MMPC and medium/long-term prediction in the MPO. The persistent predictor is applied on the de-seasonalised data which are built with 1-day difference (i.e., 96-th or 24-th sample difference, depending on 15-minute or 1-hour sampling time, respectively) on the original data. Then, the considered output of the system, the PCC power, can be expressed as follows:

$$\hat{Y} = S_f U_f + S_{fd} D_1 \tag{6}$$

where $S_f$ and $S_{fd}$ are suitable coefficient matrices; $Y = \begin{bmatrix} P_{k+1}^g \dots P_{k+N_M}^g \end{bmatrix}^T$ is the vector of future outputs and $\hat{Y} = \begin{bmatrix} \hat{y}_{k,1} \dots \hat{y}_{k,N_M} \end{bmatrix}^T$ the corresponding vector of predictions performed at time $k$; $U_f = \begin{bmatrix} u_{k,1}{}^T \dots u_{k,N_M}{}^T \end{bmatrix}^T$ is the vector of future control actions planned at time $k$, and $D_1 = \begin{bmatrix} d_{k,1} \dots d_{k,N_M} \end{bmatrix}^T$ the vector of (predicted at time k) future disturbances in the microgrid where $d_{k,i} = [P_{k,i}^l \ P_{k,i}^R]^T, i = 1, 2, \dots, N_M$.

### C. Microgrid MPC (MMPC) cost function terms and constraints

### C1. Microgrid MPC cost function terms

As will be discussed in the following sections, the energy management optimization problem can be formulated in the standard quadratic programming form:

$$\min J = U_f{}^T A U_f + B U_f + C, \text{ subject to: } D_2 U_f \leq b \tag{7}$$

where $A, B, C, D_2$ and $b$ are the suitable matrices. The considered cost function includes several additive terms spanning for the whole prediction horizon, which address different objectives as follows:

$$J_{MMPC} = w_C J^C + w_R J^R + w_P J^P \tag{8}$$

where $w_C$, $w_R$ and $w_P$ are weight coefficients and $J^C, J^R$ and $J^P$ are cost terms that account for the economic benefit resulted from electricity trading, ESS power smoothing and reference power tracking penalty. All the terms should be expressed as functions of the decision variable. The standard quadratic programming optimizations are solved using quadprog function in MATLAB which is based on interior-point-convex method with 5000 and 0.001 as maximum number of iteration and tolerance on the constraint violation allowed.

### Battery power smoothing ($J^R$)

To have a smooth behavior in ESS charging and discharging, a quadratic term of battery power is introduced as follows:

$$J^R = \sum_{i=1}^{N_M} \left( P_{k,i}^b - P_{k,i-1}^b \right)^2 \tag{9}$$

### Economic benefit ($J^C$)

This term refers to the economic benefit resulting from electricity trading with the grid and selling to the load side (disregarding monetary penalties). The power flowing through microgrid level equals to ($P_{k,i}^b - P_{k,i}^R$) which can be represented as ($P_{k,i}^g - P_{k,i}^l$) based on the power balance equation (Eq. 1). The economic benefit $J^C$ is then defined as follows:

$$J^C = \sum_{i=1}^{N_M} c_{k,i} \left( P_{k,i}^b - P_{k,i}^R \right) T_{k,i}^s \ ; \ c_{k,i} = \begin{cases} c_{k,i}^b & if (P_{k,i}^b - P_{k,i}^R) \geq 0 \\ c_{k,i}^s & if (P_{k,i}^b - P_{k,i}^R) < 0 \end{cases} \tag{10}$$

where $c_{k,i}$ is the trading electricity tariff between microgrid and the grid; $T_{k,i}^s$ is the duration of the *i-th* 1-hour block of the prediction horizon. The *if-then* condition in the cost function term requires the use of standard MIQP. Similar procedure as shown in the section III. B. is applied for $J^C$, introducing new auxiliary $C_{k,i}^C$ and binary variables $\delta_{k,i}^C$ for time instant $k+i$ during the prediction horizon as follows:

$$(P_{k,i}^b - P_{k,i}^R) \geq 0 \leftrightarrow \delta_{k,i}^C = 1 \tag{11}$$

$$C_{k,i}^C = \begin{cases} c_{k,i}^b(P_{k,i}^b - P_{k,i}^R)T_{k,i}^s & if \ \delta_{k,i}^C = 1 \\ c_{k,i}^s(P_{k,i}^b - P_{k,i}^R)T_{k,i}^s & otherwise \end{cases} \tag{12}$$

Then, the following constraints are added to the MIQP problem:

$$F_1^C \delta^C + F_2^C C^C \leq F_3^C (P^b - P^R) + F_4^C \tag{13}$$

where $\delta^C$, $C^C$, $P^b$, $P^R$ are vectors of control variables $\delta_{k,i}^C$, $C_{k,i}^C$, $P_{k,i}^b$, and estimation of renewable energy $P_{k,i}^R$, respectively.

*Tracking the reference power* $(J^P)$

Since the MPO submits the reference power to the grid operator, this power profile becomes a target for tracking of the power flows through PCC. Any violation of this committed power profile will be penalized by the grid, as described in the following formulation where PCC power $(P_{k,i}^g)$ can be rewritten as a function of $P_{k,i}^b$, $P_{k,i}^R$ and $P_{k,i}^l$ according to Eq. 1:

$$J^P = \sum_{i=1}^{N_M} PEN_{k,i} \left(P_{k,i}^g - P_{k,i}^{g,sp}\right) T_{k,i}^s = \sum_{i=1}^{N_M} PEN_{k,i} \left(P_{k,i}^b - P_{k,i}^R + P_{k,i}^l - P_{k,i}^{g,sp}\right) T_{k,i}^s$$

$$PEN_{k,i} = \begin{cases} PEN & if \left(P_{k,i}^b - P_{k,i}^R + P_{k,i}^l - P_{k,i}^{g,sp}\right) \geq 0 \\ -PEN & if \left(P_{k,i}^b - P_{k,i}^R + P_{k,i}^l - P_{k,i}^{g,sp}\right) < 0 \end{cases} \tag{14}$$

where $PEN_{k,i}$ is penalty tariff which is defined in the above formulation; $P_{k,i}^{g,sp}$ is a setpoint for the PCC power $P_{k,i}^g$ to track. At time $k$, denote $P_{k,i}^{ref}$ as the reference power at time instant $k+i$ and $P_{k,i}^{exe}$ as the average of power already flowed through PCC within the *i-th* 1-hour block of the prediction horizon, then the $P_{k,i}^{g,sp}$ can be calculated as follows:

$$P_{k,i}^{g,sp} = \frac{P_{k,i}^{ref} - P_{k,i}^{g,exe}\left(1 - T_{k,i}^s\right)}{T_{k,i}^s} \tag{15}$$

It is clear that the $P_{k,i}^{g,exe}$ takes zero value for every $i = 2, \dots, N_m$ as there is no power flowed through PCC during the 1-hour blocks following the first one, and so $P_{k,i}^{g,sp} = P_{k,i}^{ref}$ as a consequence.

Following the procedure discussed in the section III. B. , for every $i$, additive auxiliary variables $R_{k,i}^P$ and binary variables $\delta_{k,i}^P$ are defined as follows:

$$\left(P_{k,i}^b - P_{k,i}^R + P_{k,i}^l - P_{k,i}^{g,sp}\right) \geq 0 \leftrightarrow \delta_{k,i}^P = 1 \tag{16}$$

and

$$R_{k,i}^P = \begin{cases} PEN\left(P_{k,i}^b - P_{k,i}^R + P_{k,i}^l - P_{k,i}^{g,sp}\right)T_{k,i}^s & if \ \delta_{k,i}^P = 1 \\ -PEN\left(P_{k,i}^b - P_{k,i}^R + P_{k,i}^l - P_{k,i}^{g,sp}\right)T_{k,i}^s & otherwise \end{cases} \tag{17}$$

Then, the following constraints are added to the MIQP problem:

$$F_1^P \delta^P + F_2^P R^P \leq F_3^P (P^b - P^R + P^l - P^{g,sp}) + F_4^P \tag{18}$$

where $\delta^P$, $R^P$, and $P^b$ are vectors of all control variables $\delta_{k,i}^C$, $R_{k,i}^P$, and $P_{k,i}^b$, respectively; $P^R$ and $P^l$ are vectors of disturbances $P_{k,i}^R$ and $P_{k,i}^l$ (i.e, load power planned from previous time step); $P^{g,sp}$ is a vector of all PCC power setpoints $P_{k,i}^{g,sp}$ computed by Eq. 15 in the predicton horizon.

Detailed descriptions of $F_1^C$, $F_2^C$, $F_3^C$, $F_4^C$ and $F_1^P$, $F_2^P$, $F_3^P$, $F_4^P$ can be easily derived from what were illustrated above, and will not be reported here due to space limitation.

At time $k$, if we present $P_{k,i}^{g,plan}$ as the average planned PCC power computed by the MMPC for the remaining period of the *i-th* 1-hour block of the prediction horizon, then the corresponding average of the request power sent to the LC can be computed as follows:

$$\bar{R}_{k,i} = \frac{P_{k,i}^{ref} - P_{k,i}^{g,exe}\left(1 - T_{k,i}^s\right) - P_{k,i}^{g,plan}T_{k,i}^s}{T_{k,i}^s} \quad (19)$$

### C2. Microgrid MPC constraints

Apart from the introduced binary constraints (i.e., constraints on Eq. 5, 13 and 18), at every time instant k the following physical constraints of the system are considered in the MMPC.

*Power exchange with the grid:*

$$P_{min}^g \le P_{k,i}^g \le P_{max}^g, \forall\, i = 1, \dots, N_M \quad (20)$$

*ESS charge and discharge power:*

$$P_{min}^b \le P_{k,i}^b \le P_{max}^b, \forall\, i = 1, \dots, N_M \quad (21)$$

*Maximum energy stored in the ESS:*

$$E_{min} \le E_{k,i}^b \le E_{max}, \forall\, i = 1, \dots, N_M \quad (22)$$

At the end, all the discussed cost function terms and constraints will be transformed to the standard quadratic form of Eq. 7 where the decision variable $U_f$ of the MMPC contains not only the decision variables of the ESS power $P^b$, but also the auxiliary variables $C^C$, $R^P$, $z^b$ and binary variables $\delta^C$, $\delta^P$, $\delta^b$. At time k, the vector of the decision variables for time instant k+i is described as:

$$u_{k,i} = \begin{bmatrix} P_{k,i}^b & z_{k,i}^b & C_{k,i}^C & R_{k,i}^P & \delta_{k,i}^b & \delta_{k,i}^C & \delta_{k,i}^P \end{bmatrix}^T \quad (23)$$

For example, we can immediately realize the transformation of Eq. 9 into the quadratic form of Eq. 7, upon observing that $P_k^b$ is known, while all the other variables belong to $U_f$. More specifically,

$$J^R = U_f^T A_R U_f + B_R U_f + C_R \quad (24)$$

where $C_R = P_k^{b^2}$, $B_R = -2P_k^b F_0^R$, $F_0^R$ being a projection matrix such that $F_0^R U_f = P_{k,1}^b$, and $A_R = \sum_{i=1}^N E_{k,i}^R{}^T E_{k,i}^R$, $E_{k,i}^R$ being a projection matrix such that $E_{k,1}^R U_f = [P_{k,1}^b]$ and $E_{k,i}^R U_f = [P_{k,i}^b - P_{k,i-1}^b]$ $\forall i = 2, 3, \dots, N_M$. Similar manipulations can be applied to others cost terms and constraints. In the end, the optimization problem is formulated as MIQP as $U_f$ also contains binary variables.

### D. Microgrid planning optimizer (MPO)

Previous sections describe the design of microgrid MPC in which tracking reference power ($P^{ref}$) is one of their main focuses. In this section, an additive level on top of the MMPC (i.e., Microgrid planning optimizer) is studied to provide an optimal reference power $P^{ref}$. While the model and system constraints of the MPO are inherited from the MMPC, only the ESS smoothing and economic benefit cost function terms are considered in the optimization problem of this MPC as follows:

$$J_{MPO} = w_{C,O}J^{C,O} + w_{R,O}J^{R,O} \tag{25}$$

where $w_{C,O}$ and $w_{R,O}$ are weight coefficients and $J^{C,O}$, $J^{R,O}$ are cost terms that account for the economic benefit resulted from electricity trading and ESS power smoothing.

The construction of this cost function is the same as the one for the MMPC. However, notice that a longer prediction horizon $N_{MPO}$ and a longer sampling time (i.e., 1 hour) are considered in this optimizer.

## IV. Design of Load Coordinator (LC) and Building Controllers (BMPC)

In this paper, residential buildings equipped with heating systems are considered and they are managed by the BMPCs to satisfy user comfort (temperature) and economic purposes. This section starts with the general scheme of the distributed algorithm, and then the development of the BMPC in the buildings will be studied.

### A. Nomenclature

The main parameters and decision variables used in this section are described in Table II, where, for simplicity, the subscript $i$ is omitted when referring to the $i$-th step ahead in the prediction (or control) horizon.

Table II. Main nomenclature in the building level

| | |
|---|---|
| $k$ | Discrete time step based on 15-minutes sampling time |
| $m$ | Number of users involving to the building levels |
| $N_B$ | Prediction and control horizon of the BMPC |
| $T^{s,B}$ | Sampling time of the building level [Hour] |
| $I$ | Benefit tariff from changing planned consumption for users (cents/Wh) |
| $P_k^{u,j}$, | Power consumption of the building $j$ [W] |
| $dP_k^{u,j}$, $dP_{k,-}^{u,j}$ | The change of planned power consumption of the building $j$ and the change of planned power consumption for all buildings except building $j$ at time $k$, respectively [W] |
| $\overline{dP^{u,j}}$, $\overline{dP_{-}^{u,j}}$ | Setpoints computed by LC for $dP_k^{u,j}$ and $dP_{k,-}^{u,j}$, respectively [W] |
| $\bar{R}_k$ | Request for power changing from microgrid level to building level for time instant $k$ [W] |
| $T_k^j, \bar{T}_k^{\,j}$ | Temperature and temperature setpoint of building $j$ at time $k$ |
| $T_{max}^j T_{min}^j$ | Upper and lower bounds on the building $j$'s temperature respectively |
| $P_{max}^{u,j}, P_{min}^{u,j}$ | Upper and lower bounds on power consumption of the building $j$, respectively, |

### B. General scheme of the distributed algorithm

This section considers a general scenario of a network of m users (or better, "agents") that communicate to a central coordination unit to cooperatively solve a problem of sharing resources (e.g., ESS, RES or the request in the context of this paper) among them. The original form of the optimization problem of user j is presented as follows:

$$P_j: min_{x_j \in R^{n_j}} f_j(x_j); \text{ subject to } x_j \in X_j \text{ and } (x_1, x_2, \dots, x_m) \in X_c \tag{26}$$

where $x_j \in \mathbb{R}^{n_j}$ represents a vector of nj decision variables of user j and $\mathbb{R}$ denotes the set of real numbers. For each user j, $f_j(.): \mathbb{R}^{n_j} \to \mathbb{R}$, $X_j \in \mathbb{R}^{n_j}$ and $X_c \in R^{n_c}$ are objective function, constraint set of user j and the set of coupled constraints between the control variables of the users. The fuction $f_j(.)$ is convex, and $X_j$ and $X_c$ are closed convex sets for each $j = 1, 2, …, m$. To solve this optimization problem, a distributed algorithm with coordinator is here proposed based on the incremental proximal method. Specifically, in this algorithm, for each user j an additive cost function term is integrated to $f_j(.)$, in order to motivate the agreement among users. The presence of the additive cost function term causes the enlargement of the control variable sets and the change of the constraint sets of individual optimization problem $P_j$. Therefore, the set of control variables $x_j$ of the $j$-th user consists of coupled decision variables (i.e. those linked to coupled constraints and denoted as $x_{j,c}$), uncoupled decision variables (denoted as $x_{j,u}$) and the sum of coupled decision variables of all users except user j (denoted as $x_{j,c-}$) as a new decision variable. Subsequently, the coupled constraints Xc in Eq. 26 will be presented in the form $(x_{j,c}, x_{j,c-}) \in X_j$ (i.e., the local constraint sets) for every users j. The pseudo–code of the algorithm is given in Algorithm 1 as below where $\bar{x}_{j,c}$ and $\bar{x}_{j,c-}$ are denoted as setpoints computed by the LC for the coupled variables of user j and the sum of coupled decision variables of all users except user j and $\bar{x}_c$ is collection of $\bar{x}_{j,c}$ for all users.

| **Algorithm 1**: Distributed algorithm (General scheme) |
| --- |
| 1. **Initialization** |
| 2. **Set** t = 0 |
| 3.(BMPC) $x_j^{(0)} = argmin_{x_j \in X_j} f_j(x_j)$ for all users |
| 4. **Set** $\bar{x}_{j,c}^{(0)} = x_{j,c}^{(0)}$; $\bar{x}_{j,c-}^{(0)} = \sum_{h=1,h \neq j}^{m} \bar{x}_{h,c}^{(0)}$ |
| 5. **For** t = 1: L **repeat until satisfying (\*).** |
| 6.(BMPC) $x_j^{(t)} = argmin_{x_{j,c}, x_{j,u}, x_{j,c-} \in X_j} f_j(x_j) + c^{(t)}(\left\| x_{j,c} - \bar{x}_{j,c}^{(t-1)} \right\|_2^2 + \left\| x_{j,c-} - \bar{x}_{j,c-}^{(t-1)} \right\|_2^2)$ for all users |
| 7. **If** $|x_{j,c}^{(t)}(n) - \bar{x}_{j,c}^{(t-1)}(n)| < \epsilon(n)$ and $|x_{j,c-}^{(t)}(n) - \bar{x}_{j,c-}^{(t-1)}(n)| < \epsilon(n) \rightarrow (*)$ is satisfied |
| 8.**Otherwise**, |
| (LC) $\bar{x}_c^{(t)} = argmin_{\bar{x}_{1,c}, \bar{x}_{2,c}, …, \bar{x}_{m,c}} (\sum_{j=1}^{m} Q_j \left\| \bar{x}_{j,c} - x_{j,c}^{(t)} \right\|_2^2 + \sum_{j=1}^{m} Q_j \left\| \sum_{h=1,h \neq j}^{m} \bar{x}_{h,c} - x_{j,c-}^{(t)} \right\|_2^2)$ |
| 9.(LC) $\bar{x}_{j,c-}^{(t)} = \sum_{h=1,h \neq j}^{m} \bar{x}_{h,c}^{(t)}$ |
| 10. $t \leftarrow t + 1$, go to step 5 |

Initially, each user $j$ declares its desired solution $x_j^{(0)}$ by solving its original cost function in step 3. An initial value of $x_{j,c}^{(0)}$ and $x_{j,c-}^{(0)}$ are chosen as in step 4. Two different stopping criteria are presented in step 5 and step 7 where the first one is based on the maximum number of iterations and the later one is based on absoluted value of the difference in the solutions $x_{j,c}$ and $x_{j,c-}$ computed by the users at iteration t with respect to the corresponding setpoints $\bar{x}_{j,c}$ and $\bar{x}_{j,c-}$ computed by LC at iteration ($t$-1). The index n in the algorithm (step 7) presents the $n$-th element of the related vectors $x_{j,c}^{(t)}$, $x_{j,c-}^{(t)}$, $\bar{x}_{j,c}^{(t-1)}$, $\bar{x}_{j,c-}^{(t-1)}$ and $\epsilon$ (i.e., the threshold of the second stopping criteria); $h$ in steps 4, 8, 9 is the running index which indicates the users. The term $c^{(t)}$ in step 6 highlights the importance of the consensus term in the objective function. As $t$ increases, the role of the consensus term is being weighted higher to motivate the agreement among the users. An increasing function with respect to $t$ is employed for $c^{(t)}$. At every iteration t, the same value of $c^{(t)}$ is used for all users (or BMPC) in the synchronous setting of the distributed algorithm. A trivial option for $c^{(t)}$ could be $c^{(t)} = \alpha . t$, where $\alpha$ is a positive constant number which plays as an extra degree of freedom to modify the step size. $Q_j$ stands for the weight associated to the solution on the decision variables $x_{j,c}$ and $x_{j,c-}$ of user j in the LC, if a higher weight is imposed, then the solution of the corresponding user is fulfilled more strictly in the unconstrained optimization problem (step 8).

## C. Building MPC (BMPC) cost function terms and constraints

As mentioned before, every building is endowed with two controllers: one is used in the negotiation with other users and works with 15-minute sampling time (this is the BMPC), and the other one is for comfort enforcement (e.g., temperature control in this paper context) working with a much smaller sampling time, trying to track the decision taken by the negotiation unit. The second one is neglected in this paper as discussed in section II. To assess the performance of the proposed approach in the building level, the following s-domain transfer function $G(s)$ from the heating power (input) to building temperature (output) has been employed:

$$G_1(s) = \frac{4}{1 + 24500}$$ (27)

The maximum power of the building heating system is 3 kW and the gain of the model is 4°C/kW. The model is derived around an equilibrium point (12°C, 0 kW).

Denote the decision variable of building $j$ for time $k$ as $u_k^j$, $N_B$ is the prediction and control horizon of the BMPC, $\hat{Y}^j = \begin{bmatrix} T_{k,1}^j \dots T_{k,N_B}^j \end{bmatrix}^T$ is the vector of estimated building temperature of user j. Note that the value of NB is changing following the dynamic shrinking horizon depicted in Fig. 3. Let also $U_f^j = \begin{bmatrix} u_{k,1}^{j\ T} \dots u_{k,N_B}^{j\ T} \end{bmatrix}^T$ be the vector of future control actions. Then, one can express the vector of output prediction $\hat{Y}^j$ as:

$$\hat{Y}^j = S_f^j U_f^j$$ (28)

where $S_f^j$ is a suitable coefficient matrix which is derived from the building thermal model. Defining decision vector as $u_k^j = \begin{bmatrix} P_k^{u,j} & dP_k^{u,j} & dP_{k,-}^{u,j} \end{bmatrix}^T$. Refer to the algorithm 1, $P_k^{u,j}$, $dP_k^{u,j}$ and $dP_{k,-}^{u,j}$ play roles of uncoupled, coupled and the sum of coupled decision variables of all users except user $j$ variables, respectively. As opposed to the MMPC, the BMPC optimization problem is formulated in the form of standard Quadratic Programming (QP) where binary variables are not presented. Moreover, the discrete time step $k$ is here considered base on 15-minute basic.

### C1. Building MPC cost function terms

The cost function of the BMPC for each building $j$ includes several terms, that address different objectives:

$$J^j = w_T^j J^{T,j} + w_M^j J^{M,j} + w_B^j J^{B,j} + w_A^j J^{A,j}$$ (29)

where $w_T^j$, $w_M^j$, $w_B^j$ and $w_A^j$ are weight coefficients and $J^{M,j}$, $J^{T,j}$, $J^{B,j}$ and $J^{A,j}$ are cost terms that account for temperature set-point tracking, electricity consumption cost, benefit from changing planned consumption and consensus terms associated to building $j$ for reaching an agreement; while $w_T^j$, $w_M^j$, $w_B^j$ take constant values, $w_A^j$ instead is an increasing term during the user negotiation phase which equals to $c^{(t)}$ presented in the previous subsection. According to algorithm 1, the general formulation of the $J^j$ is the one in step 3 and step 6 of the algorithm. In fact, the first three terms in the cost function $J^j$ present the original local objective function of each user, while the last term is related to the additive cost function term in the algorithm. All the cost function terms should be expressed as function of the decision variables $U_f^j$ of building $j$'s in the form of Eq. 7.

*Temperature set-point tracking ($J^T$)*

A quadratic term is considered to motivate the predicted temperature $T_{k,i}^{j}$ to stay close to temperature setpoint $\bar{T}_{k,i}^{\ \ j}$. In our model, a constant setpoint is employed. However, any variation of the setpoint will not change the considered formulation.

$$J^{T,j} = \sum_{i=1}^{N_B} \left( T_{k,i}^{j} - \bar{T}_{k,i}^{\ \ j} \right)^2 \tag{30}$$

*Electricity consumption cost ($J^M$)*

The electricity consumption cost of a building is trivially computed as the product of the electricity tariff, sampling time and the consumed power ($P_{k,i}^{u,j}$) which is an element of the control variable $U_f^{j}$. Remind that in case of purchasing electricity from the grid or from the microgrid level, the same tariff of purchasing ($c^b$) is applied. Therefore, over $N_B$ step prediction horizon, the electricity cost function can be formulated as follows:

$$J^{M,j} = \sum_{i=1}^{N_B} c_{k,i}^{b} P_{k,i}^{u,j} \, T^{s,B} \tag{31}$$

$T^{s,B}$ takes a value of 0.25 which equals to the duration of one step in the BMPC in 1-hour unit (i.e., 15 minutes). Notice that the difference between the tariff of electricity sold from the microgrid level to the building level ($c^s$) and the tariff of electricity purchased by building level from the microgrid level ($c^b$, more expensive than $c^s$) generate a mutual fund for the overall microgrid. At the end of a certain period, this fund will be distributed proportionally among the users based on their contribution in supporting the microgrid level in reducing the imbalance charge.

*Benefit from changing planned consumption ($J^B$)*

In response to the request from the MC to the building level at discrete time $k$ for $i$-th step ahead in BMPC prediction horizon (denoted as $\bar{R}_{k,i}$), each building decides to change with respect to its planned consumption committed in the previous step. Given the benefit tariff ($I$) and the power change ($dP_{k,i}^{u,j}$), the benefit obtained is described as follows:

$$J^{B,j} = \sum_{i=1}^{N_B} I. \, sign(\bar{R}_{k,i}). \, dP_{k,i}^{u,j} . T^{s,B} \tag{32}$$

*Consensus term ($J^A$)*

As discussed, this term is designed to force the BMPCs to reach an agreement. In response to the desired solutions of the BMPCs, the LC provides updated set-point $\overline{dP_{k,i}^{u,j}}$ and $\overline{dP_{k,\iota,-}^{u,j}}$ for each building $j$ to be used in the following term:

$$J^{A,j} = \sum_{i=1}^{N_B} \left( dP_{k,i}^{u,j} - \overline{dP_{k,i}^{u,j}} \right)^2 + \left( dP_{k,i-}^{u,j} - \overline{dP_{k,i-}^{u,j}} \right)^2 \tag{33}$$

## C2. Building MPC constraints

### Building temperature constraints:

Besides tracking the set-point, each building needs to consider the maximum and minimum values for its temperature as hard constraints:

$$T_{min}^{j} \leq T_{k,i}^{j} \leq T_{max}^{j} \tag{34}$$

### Constraints on heating system power:

Practically, the heating system has lower and upper power bounds for its consumption that are defined as follows:

$$P_{min}^{u,j} \leq P_{k,i}^{u,j} \leq P_{max}^{u,j} \tag{35}$$

$P_{max}^{u,j}$ takes a positive value and $P_{min}^{u,j}$ is basically equal to zero when the heating system is off.

### Change of planned power consumption constraints:

In response to the request from the MC, user $j$ sends $dP_{k,i}^{u,j}$ and $dP_{k,i,-}^{u,j}$ to the LC. Both $dP_{k,i}^{u,j}$ and $dP_{k,i,-}^{u,j}$ can take either negative or positive values which stand for decreasing or increasing with respect to their planned power consumption correspondingly. At first, these control variables need to fulfil the following equalities:

$$P_{k,i}^{u,j} = P_{k,i}^{u_{plan},j} + dP_{k,i}^{u,j}, \forall\, i = 1, \dots, N_B \tag{36}$$

where $P_{k,i}^{u_{plan},j}$ is the planned power consumption of the building $j$ for time instant $k+i$. Notice that $P_{k,i}^{u_{plan},j}$ will be updated by corresponding elements of $P_{k,i}^{u,j}$ when we move to the next time instant. To compute an initial value of the $P_{k,i}^{u_{plan},j}$, the BMPC focuses only on tracking temperature setpoint. The same consideration takes place at the first sampling time of each 1-hour block to compute an intial value of $P_{k,i}^{u_{plan},j}$ for the last 1-hour block of the prediction horizon as the prediction horizon $N_B$ of the BMPC is extended with respect to the size of $P_{k-1,i}^{u_{plan},j}$ in the previous step (refer to Fig. 3, section II).

If we represent $\bar{R} = \left[ \bar{R}_{k,1} \dots \bar{R}_{k,N_B} \right]^T$ as the vector of the request at time instant $k$ for the $N_B$ steps ahead, the following constraints are assigned to the system:

a)  In the original form of $X_c$ in Eq. 26:

$$\begin{cases} 0 \leq \sum_{j=1}^{m} dP_{k,i}^{u,j} \leq \bar{R}_{k,i}, \text{if } \bar{R}_{k,i} \geq 0 \\ 0 \geq \sum_{j=1}^{m} dP_{k,i}^{u,j} \geq \bar{R}_{k,i}, \text{if } \bar{R}_{k,i} < 0 \end{cases} \forall\, i = 1, \dots, N_B \tag{37a}$$

b)  In the form in algorithm 1

$$
\begin{cases}
0 \leq dP_{k,i}^{u,j} \leq \bar{R}_{k,i} \\
0 \leq dP_{k,i}^{u,j} + dP_{k,i,-}^{u,j} \leq \bar{R}_{k,i}
\end{cases}
\text{if } \bar{R}_{k,i} \geq 0, \qquad \forall\ i = 1, \dots, N_B
$$

$$
\begin{cases}
0 \geq dP_{k,i}^{u,j} \geq \bar{R}_{k,i} \\
0 \geq dP_{k,i}^{u,j} + dP_{k,i,-}^{u,j} \geq \bar{R}_{k,i}
\end{cases}
\text{if } \bar{R}_{k,i} < 0, \qquad \forall\ i = 1, \dots, N_B
$$

(37b)

These constraints highlight the fact that, at the end, the total change of the planned power consumption of all the building and single building should not go further than the request. Notice that the request power at each time instant within the prediction horizon of the BMPC equals average request power of the corresponding 1-hour block mentioned in Eq. 19.

In the next two sections, the main results obtained with real experimental testing and simulation are illustrated and discussed.

## V. EXPERIMENTAL RESULTS

### A. Experimental facility description

The experimental validation of the proposed control algorithm was performed at Smart RUE (Smart grid Research Unit of the Electrical and Computer Engineering school), within the ERIGrid project, financing the access to European smart grid infrastructures. In these experiments, the test facility provides a small-scale, single phase microgrid operating in both grid connected mode to the local LV (Low Volgate) grid or in islanded mode [36]. Detailed description of the laboratory components (e.g., 15-kWh lead-acid storage system, 1 kW lamps, PV panels, etc.) and SCADA (Supervisory Control And Data Acquisition) system can be found in [36] and [37].

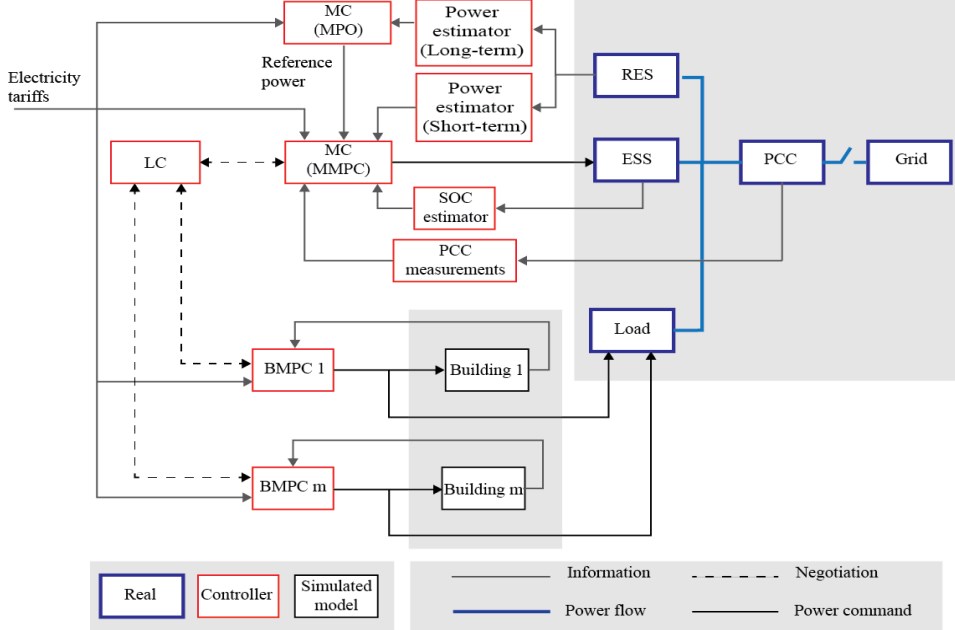

Fig. 4. Distributed control system architecture.

### B. Architecture

The overall control scheme used in the testing facilities is described in Fig. 4. While the ESS, the RES, PCC are the real equipment of Smart RUE network connected to the local LV grid, the buildings are simulated, and the related power consumption is implemented in the microgrid network using the dump load.

In the tests, the developed algorithm has been separated and executed in two different personal computers,

equipped with Generation Intel® Core™ i7 processor with 3.5GHz and 2.7 GHz frequency. In the first computer, the MMPC and the MPO are installed, while the tasks of the LC and the BMPC are executed in the second computer. To simulate the proposed distributed algorithm, the BMPCs are performed in parallel by using MATLAB (version r2017a) parallel computing toolbox in the second computer. In the end, the iteratively exchanged information between these two parts has been done through TCP/IP connection. On the other hand, UDP protocol is used to control the load and the ESS in the laboratory. For practical reasons, the load consumption in the laboratory is discretized into 10 different levels (i.e., 0, 80, 180, 280, 360, 460, 540, 640, 720, 820, 920 W), which are modulated in time in order to reach the desired values asked by the LC. Clearly, the buildings power has been scaled down to match the nominal dump load in the lab.

### C. Experimental parameters and scenarios

A set of two experiments has been carried out to investigate different aspects of the designed approach, with different control settings and under different environmental situations. Unless otherwise stated, the parameter values listed in Table III are adopted in all the experiments, where $t_{test}$ is the total period of an experiment or simulation; $N_{MPO}$, $N_M$ and $N_B$ are the prediction or control horizon of MPC in the MPO, MMPC and BMPC correspondingly.

Table III. Simulation and experiment parameters

| Parameter | Unit | Value | Parameter | Unit | Value |
|---|---|---|---|---|---|
| $P_{max}^{b,D}$ | W | 500 | $T^s$ | Minute | 15-60 |
| $P_{max}^{b,C}$ | W | 500 | $T^{s,B}$ | Minute | 15 |
| $SOC_{min}^b$ | [%] | $SOC_{ini}$-7.5 | $t_{test}$ | Minute | 240 |
| $SOC_{max}^b$ | [%] | $SOC_{ini}$+4.5 | $N_B$ | - | 9-12 |
| $C_b$ | KWh | 13.5 | $N_M$ | - | 3 |
| $\eta_c$ | [%] | 65 | $N_{MPO}$ | - | 9 |
| $\eta_d$ | [%] | 65 | m | - | 4 |
| L | | 200 or 2000 | $PEN$ | Cents /kWh | 10 |
| α | - | 0.1 | $I$ | Cents /kWh | 5 |

The value α (refer to algorithm 1), is set to 0.1 both in the experiments and the simulation, which represents a step of 0.1 in increasing the weight of coupled terms in each user cost function. In the experiments and simulations, the stop criterion of the proposed approach is set as in the algorithm 1, where the maximum number of iteration $L = 200$ or $L = 2000$ and the vector of $\epsilon$ values is chosen small enough – i.e. corresponding to 0.1% or 0.01% of the vector of the request during the prediction horizon from the microgrid level, in the experiment or simulations, respectively. The efficiencies of the battery have been manually computed at around 38.4-56.7% for both charging ($\eta_c$) and discharging ($\eta_d$). However, a value of 65% is considered for ESS efficiencies inside the microgrid coordinator, to introduce as a possible model mismatch for ESS model with respect to reality. The maximum power ESS charge and discharge power (i.e., $P_{max}^{b,D}$ and $P_{max}^{b,C}$) are also limited to 500 W which allows the ESS in the laboratory performing properly. To match the sizes of other involved components, only a portion of the ESS is employed in the experiments, the lower ($SOC_{min}^b$) and upper limits ($SOC_{max}^b$) of SOC (State of Charge) of the ESS are 4.5% and 7.5% lower or higher than the measured SOC at the beginning of any experiment (denoted as $SOC_{ini}$) respectively. The hard constraints on minimum and maximum temperature for each building (i.e., $T_{min}^j$ and $T_{max}^j$) are set equal to 1°C lower and higher than its temperature setpoint, respectively.

Regarding the cost function weights, they are chosen to balance the overall cost for both microgrid and users. Small or large values of $w_{T,j}$ determines if the buildings are benefit or comfort type. The chosen values for different weights are illustrated in Table IV.

Table IV. Cost function weights tuning in different optimizers.

| Building MPC (BMPC) | | |
|---|---|---|
| Benefit type: $w_{T,1} = 3, w_{T,2} = 6$<br>Comfort type: $w_{T,3} = 48, w_{T,4} = 96$ | $w_{M,j} = [1,1,1,1]$ | $w_{B,j} = [1,1,1,1]$ |
| Load Coordinator (LC) | | |
| $Q_j = [1,1,1,1]$ | | |
| Microgrid MPC (MMPC) | | |
| $w_P = 1$ | $w_R = 0.03$ | $w_C = 1$ |
| Microgrid Planning Optimizer (MPO) | | |
| $w_{R,O} = 0.03$ | $w_{C,O} = 1$ | |

The experiments duration was of several hours, thus catching the short-term variability of RES, high variation of the trading electricity tariff in hourly.

As mentioned, the RES power generation prediction is performed by using the persistent method. This working assumption is functional to the objective of this paper, which is not to develop novel predictor algorithms, but to assess the effect of good and bad predictions on the control architecture.

The following experiments, all involving PV panels, buildings and ESS, are performed:

1. E1. (Summer) open loop PCC Power
2. E2. (Winter) closed loop PCC Power

In the first experiment, the MMPC acts in open loop for the PCC power measurement setting, while E2 acts in closed loop setting. The difference between these schemes is on how to set $P_{k,i}^{g,exe}$ value in Eq. 15. In fact, $P_{k,i}^{g,exe}$ takes a value equal to the average of PCC power measured within the $i$-th 1-hour block in the closed loop setting, while in the open loop setting $P_{k,i}^{g,exe}$ presents the estimation value of the same factor under assumptions that there is no difference between estimated and measured values of ESS, RES and load power. Another main difference between E1 and E2 is on the period of the tests, where E1 is tested during summer, while E2 is tested during winter. For each test and experiment, the following values are computed (Table V):

1. Energy produced from RES (E_RES)
2. Prediction error for RES produced energy (ER_RES)
3. Scaled total energy consumed by users (E_L, s)
4. Energy flowed through ESS (E_B)
5. Imbalance charge for overall microgrid (I_MG)
6. Total energy cost for overall microgrid including the users on 4h basis (C_MG) including cost for energy imbalance charge

Positive values for the cost indicators represent earning and negative values represent a cost for the related entities. For the energy indicators, the users always consume energy. The RES always produces energy. The value of $ER_{RES}$ for the whole simulation period is measured based on the Root Mean Square Error (RMSE) criterion as follows:

$$ER_{RES} = \frac{100\%}{M} \frac{\sum_{k=1}^{N} \sqrt{\frac{\sum_{i=1}^{N_B}(P_{k,i}^R - P_{k+i}^R)^2}{N_B}}}{N} \tag{38}$$

where $M$ is the maximum generated power ($M = 600$W) of the RES during the test days and $N$ is the number of time steps of the whole simulation period.

### D. Experimental results

#### 1. E1. (Summer) Open Loop (OL) PCC Power:

The objective of E1 experiment is to test the overall control architecture with a PV as a source of renewable energy during summer period in a sunny day. This test is approximately 5.5 h long and started at 1:00 pm. To be consistent with the duration of the other experiments and simulation, only the results tested in the first 4 hours are reported. Random values for day-ahead promised power were assigned in this test. Sampling time of 20 minutes for $T^{s,B}$ and penalty tariff of 12 cents/kWh are also chosen.

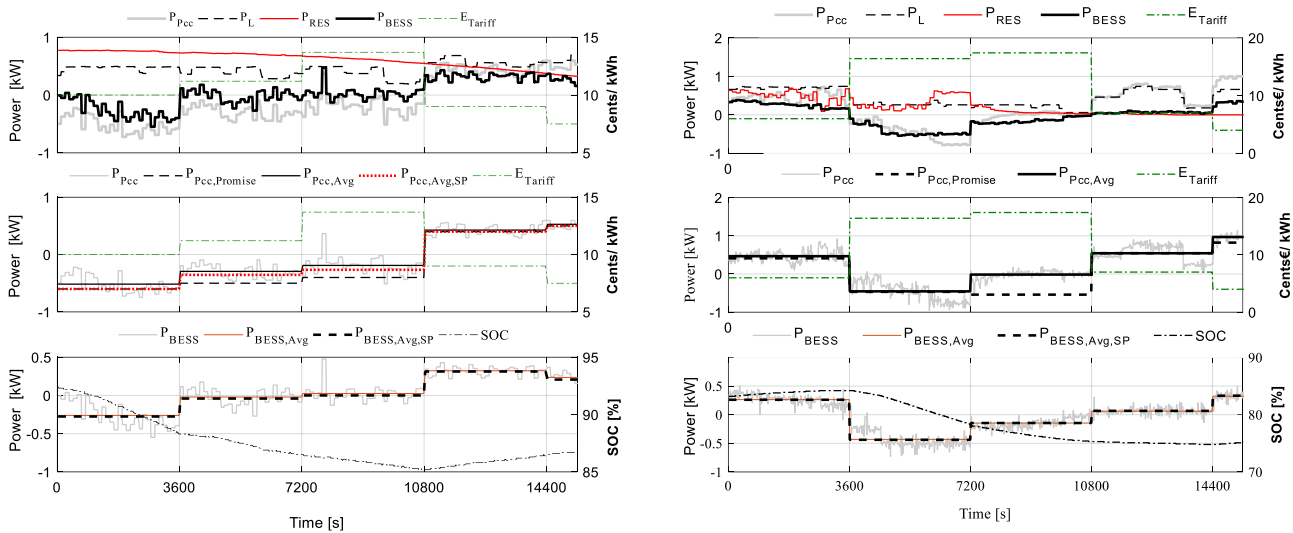

Fig. 5. Result of the first and second experiments (E1: *left figure* and E2: *right figure*) , subplot 1: power flow through all the involved components, subplot 2: PCC power (1-min average, $P_{Pcc}$), PCC power (1-hour average, $P_{Pcc, Avg}$), reference power ($P_{Pcc, Promise}$), PCC power computed ($P_{Pcc, Avg, CP}$), subplot 3: ESS power (1-min average, $P_{BESS}$), ESS power (1-hour average, $P_{BESS, Avg}$), ESS power computed ($P_{BESS, Avg, CP}$)

The temperature setpoints are 21°C for user 1 and user 3, and 21.5°C for user 2 and user 4. Users consumption profile was set to a constant value which can keep the buildings' temperature to track perfectly the temperature setpoints and this profile is known by the MMPC. MPO is not present in this test so that the reference power is randomly assigned. The initial temperature of each building is set to be equal to its setpoint.

As can be seen from the left figure of Fig. 5, the total power flow through PCC computed by our algorithm follows quite closely the reference power profile. However, the actual PCC differs by a larger quantity due to the RES prediction error and to the control error in both ESS and load side. This behavior highlighted the necessity of further investigation on the closed loop PCC power measurement which was addressed in E2.

#### 2. E2. (Winter) Closed Loop (CL) PCC Power:

Similar settings of E1 areapplied to E2 during a winter period and a cloudy erratic weather day, which caused a higher error in RES generated power prediction. This test was 4h long and started at 1:00 pm. The system is operated in closed loop with respect to PCC power. Another modification of E1 that results in E2 was the introduction of the MPO which was expected to provide a better value for the reference power than the one in E1. Different daily electricity tariffs are also tested, the variation which between two consecutive hours is higher in E2 than in E1. Notice that these electricity tariffs are used in all the simulations and all monetary items in this paper are denominated in euro (€).

Another source of prediction error is introduced by short-term PV prediction in MMPC and long/medium-term prediction in the MPO and they certainly have different quality performance in prediction. Despite the presence of various prediction and control errors, the total power flowing through PCC computed by our approach tracks very well the reference power and similar performance for the PCC actual power, since the feedback of PCC power measurement is considered in the MMPC. As shown in right figure of Fig. 5, the microgrid tends to export electricity during the peak period and import electricity during off-peak hours which brings monetary benefit. An additional experiment has been performed, to evaluate the algorithm in the case that Qj are different numbers, to represent different roles of each user within the microgrid. However, due to space limitation, this experiment results are not here reported. Regarding the power quality of the microgrid, the measured voltage and frequency of the grid and the battery inverter are measured and reported in Fig. 6. The results show an acceptable power quality during the operation of our experiment as (i) the frequency stays around 50 (Hz) with maximum deviation is 0.06 (Hz) and (ii) the voltage (RMS value) is from 222 to 234 (V) which accounts for maximum 3.5 % deviation from the nominal value of 230 (V).

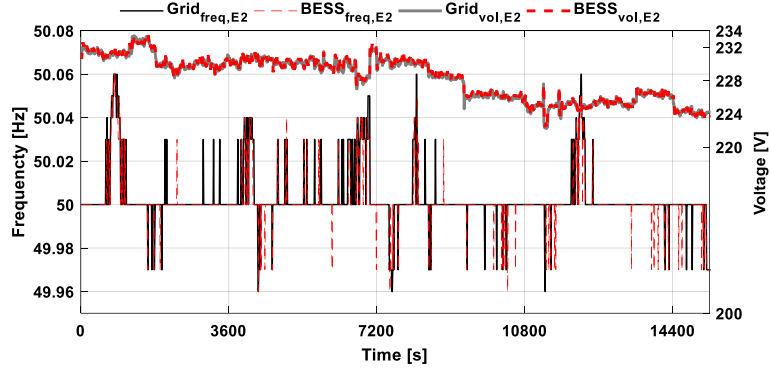

Fig. 6. Power quality in the experiment E2

Table V. Experimental and Simulation results

| ID | $E_{RES}$ | $ER_{RES}$ | $E_{L,s}$ | $E_B$ | $I_{MG}$ | $C_{MG}$ |
|---|---|---|---|---|---|---|
| Unit | Wh | % | Wh | Wh | Cent | Cent |
| Experimental Results | | | | | | |
| E1 | 2554 | 3.1 | 1801 | 65.93 | -6.32 | 0.10 |
| E2 | 937.5 | 28.9 | 1677 | -253.21 | -5.94 | -4.97 |
| Simulation Results | | | | | | |
| E2-SIM | 937.5 | 28.9 | 1675 | -234.55 | -5.56 | -3.04 |
| FLEX | 2812.5 | 0 | 4970 | -212.9 | -8.06 | -18.02 |
| FLEX1 | 2812.5 | 14.5 | 5017 | -191.2 | -11.71 | -22.98 |
| FLEX2 | 2812.5 | 28.9 | 5183 | -191.2 | -14.42 | -25.33 |
| CEN | 2812.5 | 0 | 4963 | -189.2 | -8.22 | -18.49 |
| FIX | 2812.5 | 0 | 5056 | -191.2 | -15.26 | -34.29 |

At the end, as the quality of ESS power, load control and so PCC power control are well regulated, the experimental results of E2 and its equivalent version for simulation (i.e., same parameters values, ESS, measured and predicted RES power, etc.) are expected to provide very similar if the real ESS and simulated ESS are the same. The E2-SIM is a simulation version of E2 with only difference is that the charge and discharge efficiencies of the simulated ESS (i.e., the ESS model employed as the real ESS in the simulation environment) are 50.9% and 47.1%, respectively which are derived from average performance of the real ESS. Quite similar values for most of the indicators are observed between E2 and E2-SIM in Table V. The differences can be explained by the simple simulated ESS model which can not catch some behaviors such as self-discharge, highly nonlinear SOC of the real one.

## VI. SIMULATION RESULTS

Three sets of simulations are here illustrated to point out different characteristics of the system which are the impact of the flexible load, RES generated power prediction error and comparison of the proposed control

techniques, as follows:

1. S1. Comparison between fixed and flexible load;
2. S2. Comparison between different levels of RES generated power prediction errors;
3. S3. Comparison between the proposed DMPC and centralized MPC in the building level;

The starting point of all the simulations is the second experiment (E2). Comparing to E2, the closest simulation is FLEX where the E2's system settings, parameters and RES generated power are employed in this simulation. However, the considered RES prediction herein is the perfect one which implies that the prediction and measured values are identical. On the other hand, the number of buildings involving is also change from four buildings in E2 to 12 buildings in our simulation which can be represented by replicating four building types introduced in Table IV for three different building models as follows:

$$G_1(s) = \frac{4}{1 + 24500}; \; G_2(s) = \frac{5}{1 + 25200s}; \; G_3(s) = \frac{6}{1 + 25200s} \tag{39}$$

Accordingly, the PV production is also increased three time to match someway to the level of the consumption side. However, the size of the ESS (i.e., power, capacity, SOC limits) remains unchanged to promote better the role of the flexible load in our simulations. Moreover, compared to E2, estimated ESS model is considered in the simulation environment while the real ESS is tested in E2. A better BESS with respect to the one in the experiments is employed in the simulation with the charge and discharge efficiencies of 65% in ESS model considered in MPC formulation of the MPO and the MMPC, while in the simulated ESS, 55% of the charge and discharge efficiencies is considered to provide another source of model mismatch in the simulation.

Four different alternative versions of the FLEX will be studied which are labeled as FLEX1, FLEX2, FIX and CEN. Detailed description of these simulations will be reported in the following section while the simulation results are shown in Fig. 7, Fig. 8 and Table V.

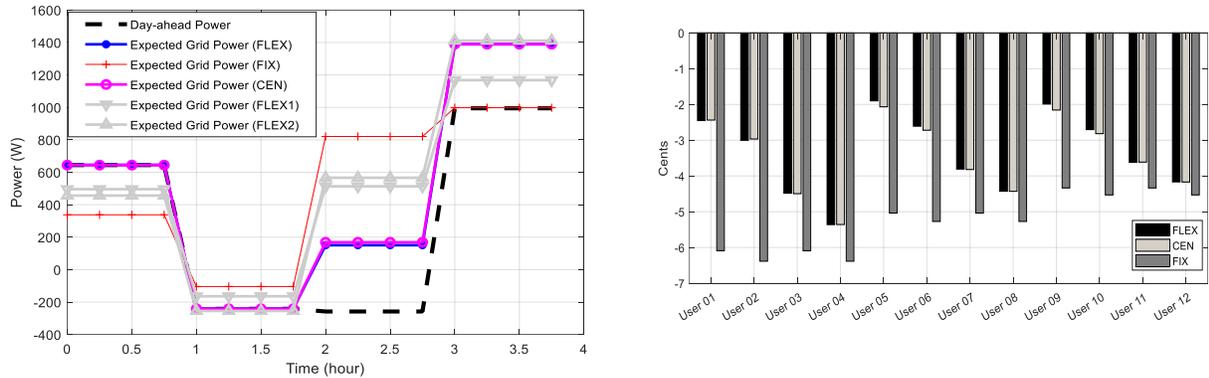

Fig. 7. Reference power tracking (*left figure*) and Net bill for all users (*right figure*) for FLEX, CEN and FIX

### 1. S1: Fixed and flexible load comparison

With respect to the FLEX, the FIX does not consider load flexibility. This simulation is the case when there is no microgrid negotiation and user negotiation phases.

As shown in Table V, the overall microgrid reduces significantly the cost from 34.29 cents in the FIX to 18.02 cents in the FLEX. Specifically, the net benefit for the microgrid level (i.e., imbalance charge, energy cost for microgrid level, incentive paid to the users) increases from 23.79 cents in the FIX to 25.37 cents in the FLEX which is equivalent to 6.6% improvement. In fact, Fig. 7 shows a better reference power tracking in the case of the FLEX instead of the FIX which will, therefore, reduce the imbalance charge for microgrid level. However, it is clear that the levels of comfort of the users in FIX (i.e., users in FIX focus only on temperature tracking) are much higher than the ones in FLEX since in FLEX the users changed their intial expected power consumption to support the overall microgrid (see Fig. 8).

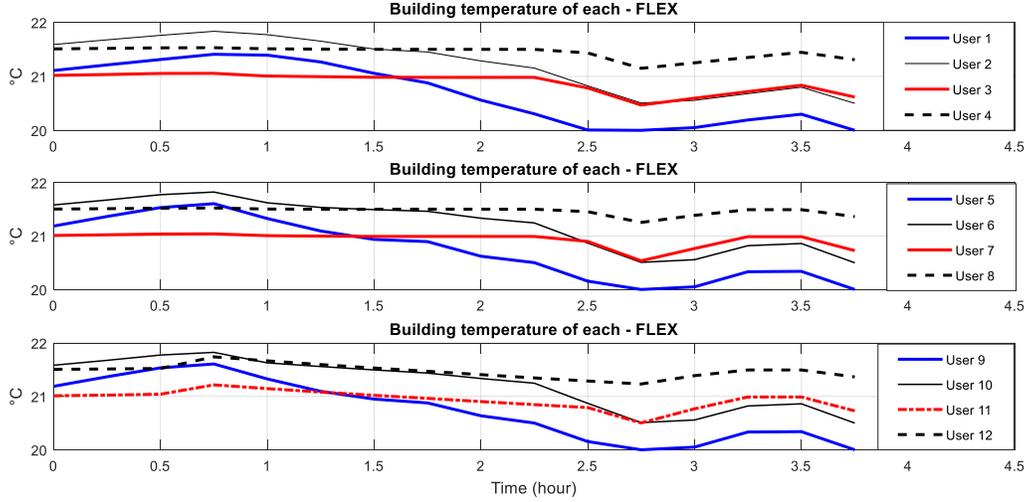

Fig. 8. Building temperature of each user in FLEX

Regarding the users' bill, their net bills are computed by subtracting the energy cost and the incentives acquired from the MC and the mutual fund; positive values represent earning and negative values represent a cost for the related entities. This information is shown in Fig. 7 which indicates a significant improvement in the net bill for all users by considering the load flexibility. For the sake of comparison, although the users in the FIX case do not contribute to reducing the imbalance charge for microgrid level, the mutual fund is still distributed equally to each of them. Clearly, significant improvement in the net bills for benefit type users with respect to the comfort type users is observed Fig. 7. Regarding the building models, the users of the first building model (i.e., user 1, 2, 3 and 4) gain more money than the users of second and third building models (i.e., user 5, 6, 7, 8 and user 9, 10, 11, 12 respectively) as the first building model has less gain and longer time constants which allows the buildings of this type change their expected power profile with less change in their comfort.

### 2. S2: Different levels of RES generated power prediction error comparison

This section is devoted to make a comparison between different versions of the FLEX where the levels of prediction error ranges from 0% (i.e. perfect short-term prediction) in the FLEX to 28.9% (the short-term prediction used in E2 during the experiment) in FLEX2. An intermediate prediction error level 14.5% in FLEX1 whose predictions are manipulated by mixing the prediction values of FLEX2 and perfect prediction values in FLEX. The results are summarized in Table V that shows a significant improvement in the cost for the overall microgrid when improving the quality of the prediction. Better planning for the ESS activities and the requested power to the building level are obvious reasons for this improvement. As a matter of fact, a large difference in terms of the overall microgrid cost is observed between the worst (i.e., ERRES=28.9%) and the best case (i.e., ERRES=0%). A part of this difference comes from the quality of the reference power tracking in these simulations as shown in Fig. 7 which results in the imbalance charge paid to the main grid being 8.06 cents, 11.71 cents, and 14.42 cents for FLEX, FLEX1, and FLEX2, respectively.

### 3. S3: Distributed MPC with coordinator and centralized MPC

This simulation focuses on evaluating the results obtained from the proposed distributed algorithm with respect to the centralized MPC result which is considered as an optimal solution for the system. Comparing to the FLEX, the CEN replaces the multiple BMPCs by a single centralized MPC (CMPC). Consequently, the need of receiving the setpoint from the LC to CMPC is not required, while the models and settings of all buildings are revealed for the CMPC. Consider the single centralized optimization in the form of:

$$P: min_{x \in R^n} \sum_{j=1}^{m} Q_j^* f_j(x)$$

(40)

$$\text{subject to } x \in X_c \cap (\cap_{j=1}^{m} X_j)$$

where $x$ presents all the original control variables (i.e., $x_{j,u}$ and $x_{j,c}$) of the involved users. $Q_j^*$ are weights associated to each agent local cost function in the centralized optimization problem. We tested in simulation the case when all values of $Q_j$ in the central coordination unit of the LC are equal to 1 which is expect to provide similar solution as solving optimization problem defined in Eq. 40 where all $Q_j^*$ equal to 1. Indeed, the reported results in Table V, Fig. 7 indicate the similarities between the solutions of the distributed and centralized approaches. Fig. 9 shows an example of a set of trajectories of difference of variables $dP_{k,i}^{u,j}$ in FLEX and CEN for all user $j$ with respect to the number of iterations for a fixed number of $k$ and $i$.

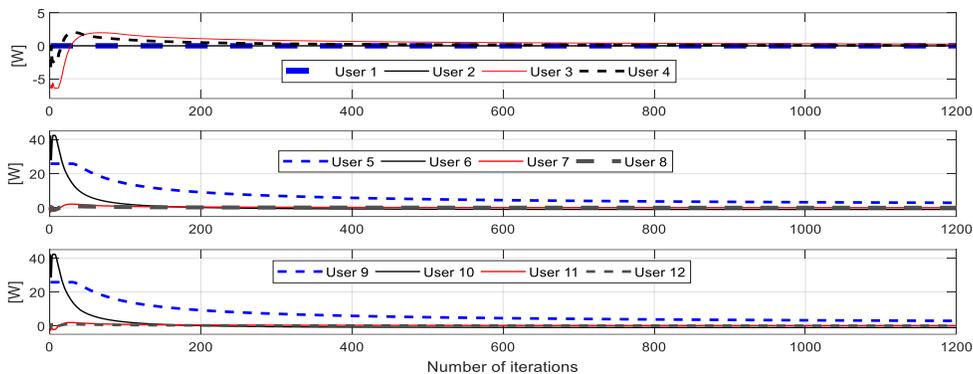

Fig. 9. An example on the difference between solutions from distributed and centralized MPC

As the number of iterations increases the difference between the solutions from distributed and centralized MPC algorithm gets smaller. In detail, the maximum difference between couples of solutions (i.e., User $j$ – FLEX and User $j$ – CEN for all j) reduces from 6.6 W and then to 2.91 W if $\epsilon$ is set to 0.01% and 0.001% which causes the negotiation phase to stop at the 333rd, and the 1218th iteration correspondingly. In the former case, the total duration for completing all the negotiation phases in this specific time instant is 132s. Notice that the main part of the time duration (more than 94%) is for computing optimization solution in BMPCs. This suggests that if the BMPCs optimization problem is solved in a distributed way in different controllers, then the duration of the negotiation phase will be significantly decreased.

## VII. CONCLUSIONS

This paper proposes an innovative two-layer hierarchical control framework for smart microgrids, integrating a distributed model predictive control framework with a coordination algorithm for smart buildings in smart microgrids. The overall framework includes day-ahead market interactions down to inter-user negotiations initiated by a request to change the energy profile in the day-ahead market. The scheme allows the user to be directly involved in any decision related to the adaptation of their profile. All the decisions are taken in an optimal way, and according to a predictive approach. In the hierarchical control framework, the microgrid level, which employs a standard MPC framework in managing the microgrid operation, is combined to the building level for further improving the overall microgrid performance by exploiting the flexibility in changing their consumption. The negotiation phase between the two hierarchical levels is updated in a sampling time shorter than the time scale of the main target factors in the microgrid operation which gives a chance to refine the decision and enhance the performance to compensate for any forecast errors. On the other hand, to reach an agreement among multiple users in the building level, an iterative, increment proximal minimization in MPC framework with the dynamic shrinking horizon is studied. The weights in the central coordination unit are introduced to provide a further tool to classify the ability of each user in reaching the consensus solution. The proposed control architecture has been implemented and tested in the laboratory Smart RUE in NTUA in Athens, with a microgrid, which consists of RES, controllable load, storage system, and grid connection. In addition to the experiments, an extensive simulation campaign has been performed in order to strengthen the analysis of the control framework and enhance the performance the control framework for any forecast error. Both the simulation and experimental results illustrate the potentials of the proposed approach, in terms of adaptation to different scenarios, convergence, and optimality with respect to centralized solutions. And the realtime experiments performed with normal laptops show that computationally the approach is feasible.

All the above stimulates further research in this field, in particular, towards the automatic construction of a day-ahead user consumption profile in a robust way, an iterative negotiation between high and low-level layer at every sampling time, in order to reduce differences with respect to the optimal centralized solution, and integration with low-level real-time control of buildings. Also, more experimental steps are envisaged, for further validation of the overall approach.